# An Atomic-Scale View of CO and $H_2$ Oxidation on a Pt-$Fe_3O_4$ Model Catalyst

Roland Bliem,[a] Jessi van der Hoeven,[b] Adam Zavodny,[c,d] Oscar Gamba,[a] Jiri Pavelec,[a] Petra de Jongh,[b] Michael Schmid,[a] Ulrike Diebold,[a] and Gareth S. Parkinson*[a]

**Abstract:** Metal-support interactions are frequently invoked to explain the enhanced catalytic activity of metal nanoparticles dispersed over reducible metal-oxide supports, yet the atomic scale mechanisms are rarely known. Here, we use scanning tunneling microscopy to study a $Pt_{1-6}/Fe_3O_4$ model catalyst exposed to CO, $H_2$, $O_2$, and mixtures thereof, at 550 K. CO extracts lattice oxygen at the cluster perimeter to form $CO_2$, creating large holes in the metal-oxide surface. $H_2$ and $O_2$ dissociate on the metal clusters and spill over onto the support. The former creates create surface hydroxyl groups, which react with the support to desorb water, while atomic oxygen reacts with Fe from the bulk to create new $Fe_3O_4$(001) islands. The presence of the Pt is crucial because it catalyses reactions that already occur on the bare iron-oxide surface, but at higher temperatures.

Variations in the CO oxidation activity of nominally similar metal nanoparticles dispersed over different metal oxides [1] are a clear indicator of metal-support interactions, but the mechanisms by which the metal oxide gets involved are difficult to ascertain. The support is known to transfer electrons to (or from) the clusters [2], modifying their shape and adsorption properties, and adsorb reactants and promoters (e.g. water) that can speed up the reaction [3]. At elevated temperature, the so-called Mars-van Krevelen (MvK) mechanism [1i] can occur, where CO molecules extract lattice oxygen ($O_{lattice}$) at the cluster perimeter forming $CO_2$, and gas phase $O_2$ repairs the surface to complete the catalytic cycle. While there is mounting evidence that MvK plays a role when reducible metal oxides are utilized as the support [1i], there is little atomic-scale information available to better understand how the process occurs.

In this paper, we use scanning tunneling microscopy (STM) to follow the evolution of a Pt/$Fe_3O_4$(001) model catalyst exposed to CO, $O_2$ and $H_2$ at 550 K. Holes and islands in the vicinity of Pt clusters provide direct evidence of reduction and oxidation of the metal oxide in CO and $O_2$ rich atmospheres respectively, while dissociation and spillover of $H_2$ leads to hydroxylation of the support lattice. We interpret our results as the metal catalyzing reactions that otherwise occur between the reactants and support at higher temperatures.

An STM image of the Pt/$Fe_3O_4$(001) model catalyst utilized as the basis for this work is shown in Figure 1. The $Fe_3O_4$(001)-support [4] exhibits large, flat terraces (Fig. 1, inset) characterized by rows of protrusions related to surface Fe atoms in STM images. The rows rotate by 90° from terrace to terrace, a consequence of the spinel structure. Surface O atoms are not imaged because they have no density of states in the vicinity of the Fermi energy [5]. Surface OH groups, formed through the reaction of water and oxygen vacancies during sample preparation [6], modify the density of states of the neighboring Fe atoms making them appear brighter in STM images [7]. The surface exhibits a ($\sqrt{2}\times\sqrt{2}$)R45° reconstruction [4], which binds metal adatoms in one particular site per unit cell to temperatures as high as 700 K [4, 8].

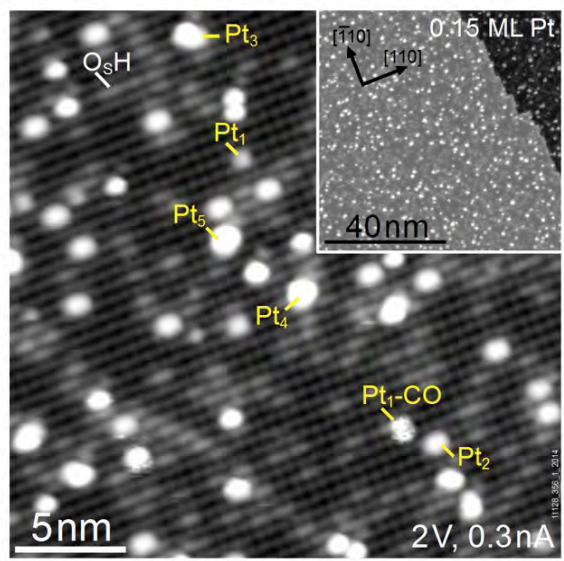

**Figure 1. The $Pt_{1-6}/Fe_3O_4$(001) model catalyst.** The $Fe_3O_4$(001) support exhibits rows of protrusions related to surface $Fe_{oct}$ atoms. Elongated protrusions on the $Fe_{oct}$ rows are surface OH groups. $Pt_{1-6}$ clusters (large, oval-shape protrusions) are formed by exposing $2.7\times10^{17}$ $m^{-2}$ $Pt_1$ adatoms to $1\times10^{-7}$ mbar CO for 10 minutes at room temperature. On adsorption of CO, otherwise stationary $Pt_1$ adatoms become mobile and agglomerate. Pt cluster sizes are assigned on the basis of STM movies in which the agglomeration was tracked atom-by-atom. (inset) On a large scale, the support exhibits large, flat terraces on which the metal particles are randomly distributed.

[a]  R. Bliem, O. Gamba, J. Pavelec, Prof. M. Schmid, Prof. U. Diebold, Ass. Prof. G. S. Parkinson
Institute for Applied Physics
TU Wien
Wiednerhauptstrasse 8-10, 1050 Wien, Austria
E-mail: Parkinson@iap.tuwien.ac.at
[b]  J. van der Hoeven, Prof. P de Jongh
Debye Institute for Nanomaterials
Utrecht University
P.O. Box 80.000, 3508 TA Utrecht, The Netherlands
[c]  A. Zavodny
Institute of Physical Engineering
Brno University of Technology, 61669 Brno, Czech Republic
[d]  A. Zavodny
CEITEC BUT
Brno University of Technology, 61669 Brno, Czech Republic

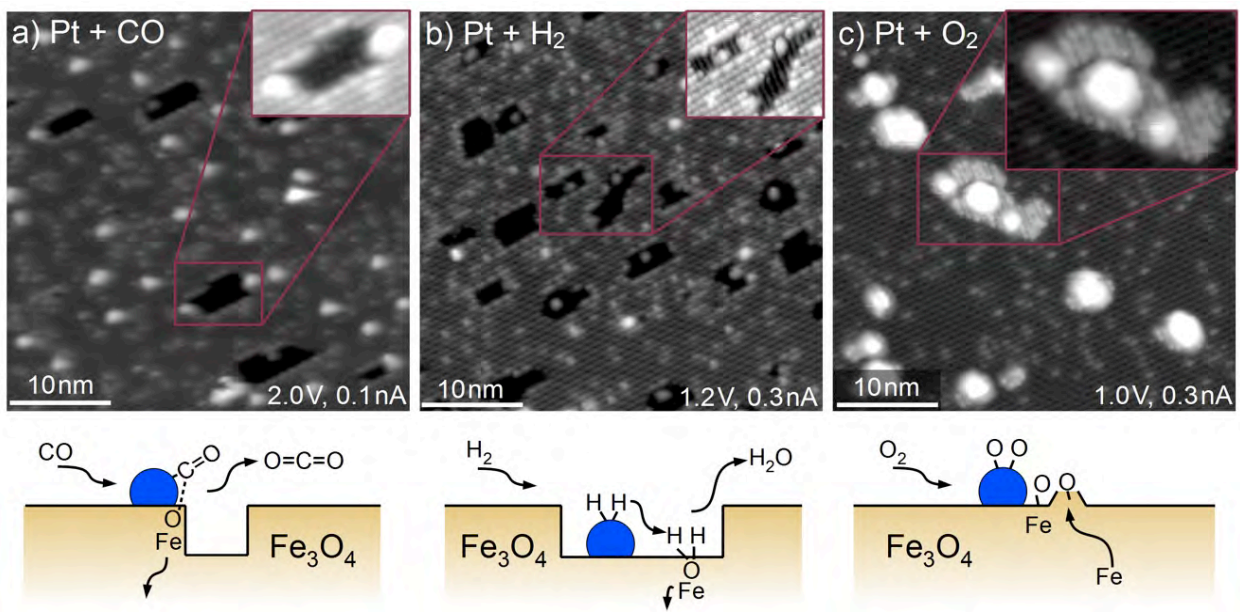

**Figure 2.** STM images acquired following exposure of the Pt/Fe$_3$O$_4$(001) model catalyst to 1×10$^{-7}$ mbar CO (a), H$_2$ (b), and O$_2$ (c) at 550 K. (a) Following exposure CO (60 minutes) approximately 50 % of the Pt clusters sit at the edge or inside monolayer holes in the Fe$_3$O$_4$(001) terrace. The (90° rotated) rows of the next Fe$_{oct}$-O layer can be seen inside the holes (inset). The sketch illustrates how CO extracts O$_{lattice}$ from cluster perimeter, and CO$_2$ desorbs from the surface. Undercoordinated Fe atoms diffuse into the Fe$_3$O$_4$ bulk. (b) Following exposure to H$_2$ (20 minutes) the clusters reside inside monolayer holes in the Fe$_3$O$_4$(001) terrace. Pt adatoms are observed at regular terrace sites, as well as an increased coverage of surface OH groups. The sketch shows how H$_2$ dissociates on the Pt cluster, spills over onto the support, and reacts with O$_{lattice}$ to desorb molecular water. (c) Following exposure to O$_2$ (20 minutes) all Pt clusters reside on top of islands with a step height of 2 Å. On the larger islands the (90° rotated) rows of the next Fe$_3$O$_4$(001) layer are clearly visible. The islands result from spillover of atomic O onto the Fe$_3$O$_4$ support, which reacts with Fe that diffuses out from the bulk.

In this work, randomly distributed Pt clusters with sizes in the range of one to six atoms (Fig. 1a) were prepared by CO-induced sintering of 2.7×10$^{17}$ Pt adatoms/m$^2$, as demonstrated previously for Pd [8f]. XPS measurements (see Fig. S3) reveal that the CO remains on the clusters, but desorbs when the sample is heated to 550 K.

Exposing the Pt$_{1-6}$/Fe$_3$O$_4$ system to CO, O$_2$, and H$_2$ at room temperature results in no discernible difference in STM (Fig. S2). However, exposure to the same gases at 550 K induces significant modification to the support morphology in the vicinity of the clusters. Following CO exposure, (Fig. 2a, $p_{CO}$=1×10$^{-7}$ mbar, 60 minutes) large holes covering 3.7 % of the surface are observed in the Fe$_3$O$_4$(001) terrace. The step height (2 Å) is consistent with the distance between Fe$_{oct}$-O planes, and the (90° rotated) rows of the underlying Fe$_{oct}$-O layer are clearly visible inside the larger holes (see inset). Each hole is associated with (at least) one Pt cluster, which typically sits at the step edge, but only half of the clusters are associated with holes. Holes are also created when the model catalyst is exposed to H$_2$ (Fig. 2b, $p_{H2}$=1×10$^{-7}$ mbar, 20 minutes), but now all clusters appear inside holes that cover 8.7 % of the surface. Since this occurs in just 20 minutes, the process is clearly more efficient with H$_2$. Some redispersion results in the observation of isolated Pt$_1$ adatoms on the surface.

None of the morphological changes described above occur in the absence of Pt (see Fig. S2), but CO and H$_2$ reduce iron oxides at higher temperatures producing CO$_2$ and water respectively [9]. In the case of CO, the Pt cluster efficiently adsorbs the molecule and delivers it to the cluster/support interface where it extracts O$_{lattice}$ to create CO$_2$. Since some clusters are associated with the removal of many tens of O$_{lattice}$ atoms, while others remain on the pristine substrate, extraction of the first O$_{lattice}$ atom most likely presents a significant energetic barrier. This is in line with recent DFT+U calculations for CeO$_2$(111), where undercoordinated O atoms at step edges were found to be more easily removed by CO than terrace oxygen atoms [10]. Since clusters are mostly found on the upper step edge of large holes, the clusters most likely diffuse along the step once a hole is nucleated, extracting more and more lattice oxygen as they go. One final question remains: Why are holes observed instead of isolated point defects? The answer is that undercoordinated Fe atoms created by the surface reduction diffuse into the bulk forming interstitials at elevated temperature, as observed previously [11[8e, 8g]]. Oxygen vacancies are not common defects in iron oxides, which handle stoichiometric variation via the cation sublattice [11].

In H$_2$, dissociation on the clusters followed by spillover creates surface OH groups, which are known to react with O$_{lattice}$ to desorb water above ≈ 500 K [6, 12]. Since OH groups can diffuse over the metal-oxide surface at 550 K [6, 12], O$_{lattice}$ extraction can occur away from the cluster. In this way, once a hole is nucleated the step edge recedes from the cluster with time, leaving the cluster in the bottom of the hole. Such autocatalytic reduction of iron oxide by hydroxyl groups has been reported previously [13]. Interestingly, since the clusters appear trapped inside the holes they create, pretreating the Pt/Fe$_3$O$_4$(001) system in H$_2$ at 550 K could be an effective strategy to mitigate thermal sintering during CO oxidation.

When the Pt$_{1-6}$/Fe$_3$O$_4$ model catalyst is exposed to O$_2$ at 550 K (Fig. 2c, p$_{O2}$=1×10$^{-7}$ mbar, 20 minutes), the Pt clusters appear atop small islands with a step height of 2 Å. The characteristic rows of Fe$_3$O$_4$(001) are clearly visible on the larger islands, suggesting that the clusters catalyze the growth of new stoichiometric $Fe_3O_4$(001) surface. A significant reduction in the density of clusters suggests gas-assisted sintering occurs in the $O_2$ atmosphere. The growth of new $Fe_3O_4$(001) surface has been observed above 900K on the clean surface [14]. Here the Pt clusters catalyze the rate-limiting step, $O_2$ dissociation, and highly reactive atomic O species spill over onto the support where they react with Fe supplied from the bulk. Thus, as well as providing a sink for excess Fe during reduction, the $Fe_3O_4$ bulk also supplies Fe for the growth of new $Fe_3O_4$(001) surface during reoxidation [14]. Similar metal-activated regrowth leads to encapsulation of the metal clusters and a loss of catalytic activity for Pt and Pd clusters on $TiO_2$ surfaces [15], but our low energy ion scattering (LEIS) measurements (Fig. S4) suggest that the Pt clusters are not encapsulated by the $Fe_3O_4$(001) islands. Zhang et al. recently observed encapsulation of larger Pt clusters to occur above 650 K in UHV [8a].

The data shown in Fig. 2 clearly show that etching and regrowth of the $Fe_3O_4$(001) support are catalyzed by Pt clusters. To test whether similar processes still occur when multiple gases are present we heated the Pt/$Fe_3O_4$ model catalyst in mixtures of CO, $O_2$, and $H_2$ at 550 K. In a 1:1 mixture of CO and $O_2$, the changes to the support morphology are significantly less than with the individual reactants (Fig. S5). This could be due to two effects: either both $O_{lattice}$ extraction and island growth act at the same time and their effect on the morphology cancel out, or the presence of one gas blocks activation of the other (poisoning). To check, we exposed the sample to CO:$O_2$ mixtures of varying composition, and found small holes begin to appear in a 50:1 CO excess, while small islands begin to appear in a 5:1 $O_2$ excess. This suggests that the etching and regrowth processes still occur when both CO and $O_2$ are present in the gas phase, although we cannot exclude that direct reaction processes also take place under such conditions. Finally, exposure of the system to a 1:1 mixture of CO and $H_2$ for 20 minutes results in holes that extend multiple layers deep into the support (Fig. S5). Such a roughening of the surface suggests that the addition of $H_2$ to the gas mixture should help activate Pt clusters for the CO oxidation reaction.

In conclusion, the holes and islands that appear in CO/$O_2$ rich environments provide clear evidence of reduction and oxidation of the $Fe_3O_4$ support being catalyzed by Pt clusters. The legacy of the reactions is strikingly clear in STM images because the $Fe_3O_4$ bulk provides a sink for Fe atoms, which readily diffuse to and from the surface at the reaction temperature. In general, our data provide an atomic-scale view into metal-support interactions that occur in heterogeneous catalysis.

## Acknowledgements

This material is based upon work supported as part of the Centre for Atomic-Level Catalyst Design, an Energy Frontier Research Centre funded by the U.S. Department of Energy, Office of Science, Office of Basic Energy Sciences under Award Number #DE-SC0001058. G.S.P. and O.G. acknowledge support from the Austrian Science Fund project number P24925-N20. R.B. acknowledges a stipend from the Vienna University of Technology and Austrian Science Fund doctoral college SOLIDS4FUN, project number W1243. U.D. and J.P. acknowledge support by the European Research Council advanced grant "OxideSurfaces". AZ acknowledges support by the European Regional Development Fund – (CEITEC - CZ.1.05/1.1.00/02.0068). The authors acknowledge Prof. Mao (Tulane University) for providing the synthetic $Fe_3O_4$ sample used in the work.

**Keywords:** Scanning probe microscopy • Supported catalysts • Oxide surfaces • Support effects • Mars van Krevelen mechanism

**Entry for the Table of Contents** (Please choose one layout)

Layout 1:

## COMMUNICATION

Scanning tunnelling microscopy measurements reveal etching and regrowth of an iron-oxide support occurs in the vicinity of Pt clusters during CO and $H_2$ oxidation. The presence of the Pt catalyses reduction/oxidation reactions that occur on the bare support at higher temperatures.

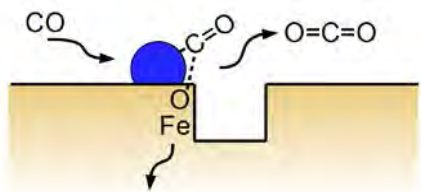

*Roland Bliem, Jessi van der Hoeven, Adam Zavodny, Oscar Gamba, Jiri Pavelec, Petra de Jongh, Michael Schmid, Ulrike Diebold, and Gareth S. Parkinson\**

*Page No. – Page No.*

**An Atomic-Scale View of CO and $H_2$ Oxidation on a Pt-$Fe_3O_4$ Model Catalyst**

This supplement contains:

- Experimental details
- Figure S1 showing STM images acquired after annealing the $Pt_{1-6}/Fe_3O_4(001)$ sample at 550 K in UHV
- Figure S2 showing STM images after annealing the pristine $Fe_3O_4(001)$ surface (without Pt) in CO, $H_2$, and $O_2$ at 550 K
- Figure S3 showing XPS spectra of three preparations of $Pt/Fe_3O_4(001)$
- Figure S4 showing low-energy ion scattering spectra for the $Pt_{1-6}/Fe_3O_4(001)$ sample after annealing to 610 K in UHV and in $O_2$
- Figure S5 showing STM images of $Pt_{1-6}/Fe_3O_4(001)$ exposed to mixtures of CO and $O_2$ in the ratios (a) 1:5, (b) 1:1, and (c) 50:1 ($p_{CO}=1\times10^{-7}$ mbar) at 550 K. In (d) a 1:1 mix of CO and $H_2$ is shown.

**Experimental details**

The experiments were performed on a synthetic $Fe_3O_4$(001) single crystal grown using the floating zone method (the sample was used for data presented in Figures 1-2, S1, S2, S5b-d) and a natural $Fe_3O_4$(001) single crystal (Figures S3, S4, S5a). Both samples were prepared by sputtering with $Ar^+$ ions ($E_{Ar}$ =1 keV, ion current density $j_{sample} \approx 6.5$ µA/cm$^2$, 10 min) and annealing in $O_2$ at 870 K ($p_{O2}$= $10^{-6}$ mbar, 10 min). Temperatures were measured with a K-type thermocouple attached to the sample holder. Up to a temperature of 600 K, the systematic uncertainty of the measurement is estimated as ±20 K for each crystal. In the annealing experiments, the sample holder was heated slowly (>10 min) by resistive heating to ensure the surface reaches equilibrium at the temperature set point. No differences between the two samples were observed at room temperature. The nominal temperature required for island/hole growth in $O_2$ and CO atmospheres differed by ≈ 60 K between the two samples, however. We attribute this difference to the different thickness and mounting of the samples, but we cannot discount that a small difference in sample stoichiometry could also be present, and may play a role. The deposition rate for Pt (≈1 ML/min, 1 ML = 1 adatom per unit cell = $1.42 \times 10^{14}$ atoms/cm$^2$) was calibrated using a temperature stabilized quartz crystal microbalance (QCM). The coverages given in the description of the data are determined by the QCM calibration, which is in line with estimates from counting adatoms before sintering. The Pt cluster sizes given in Figure 1 are derived from an analysis of STM movies where the formation of each cluster is observed atom-by-atom [1]. The LEIS data were acquired using a Specs IQE 12/38 ion source accelerating $He^+$ ions to 1025 eV and a Specs EA 10 Plus analyser; the scattering angle of the detected ions is 90°.

The XPS data were acquired in a separate UHV chamber equipped with a He flow cryostat, direct current sample heating, an Al Kα X-ray source with a Focus 500 monochromator, and a SPECS Phoibos 150 analyzer. The analyzer has been calibrated (±0.1 eV) using Au as a reference. In this UHV chamber the sample was prepared by cycles of sputtering with $Ne^+$ ions ($E_{Ne}$ =1 keV, $j_{sample} \approx 2.3$ µA/cm$^2$, 10 min) and annealing in $O_2$ (using a directional doser) at 870 K (chamber pressure $p_{O2}$ = $8 \times 10^{-8}$ mbar, 15 min).

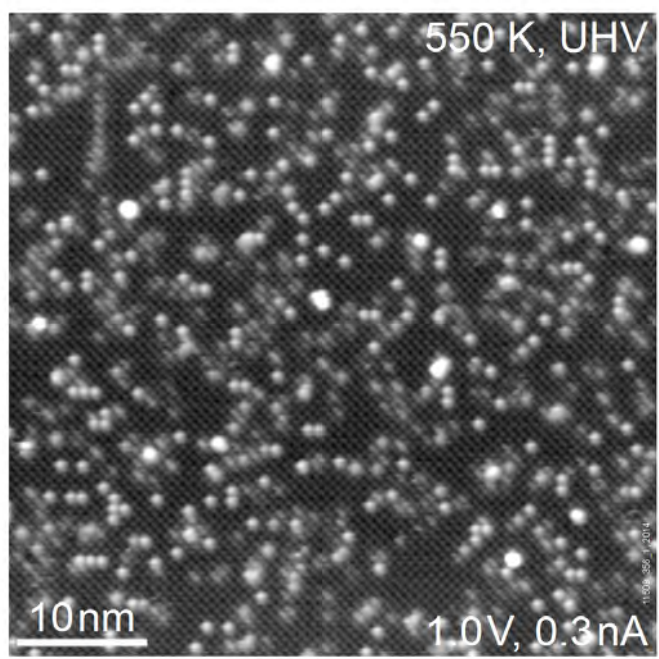

**Figure S 1: Control experiment: heating in UHV.** STM image acquired following annealing the $Pt_{1-6}/Fe_3O_4$(001) sample to 550 K in UHV. No holes or islands were observed. The adatom density is increased as compared to Figure 1, indicating redispersion of Pt caused by heating in UHV.

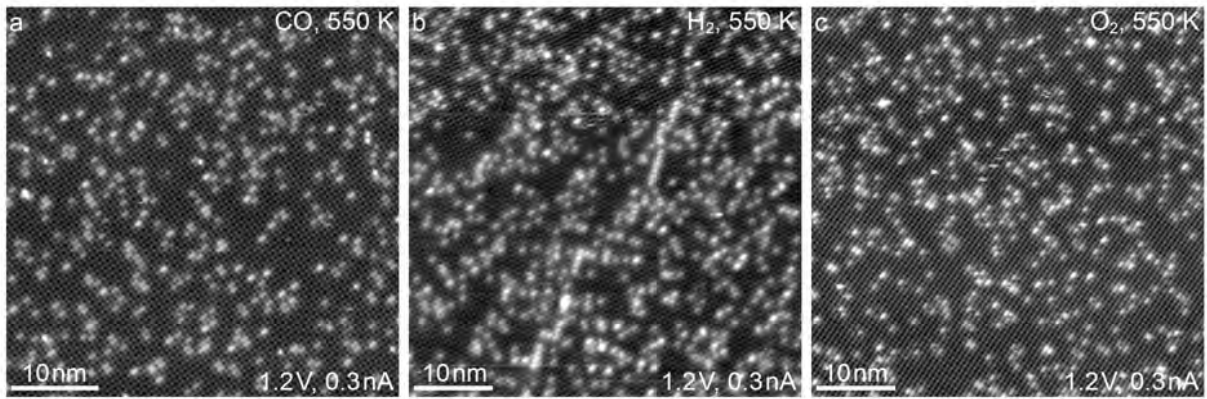

**Figure S 2: Control experiments – heating clean $Fe_3O_4$.** STM images acquired following annealing the clean $Fe_3O_4$(001) sample in gas pressures identical to the ones used in the experiments described in the main text. (a) $1\times10^{-7}$ mbar CO for 60 min at 550 K. (b) $1\times10^{-7}$ mbar $H_2$ for 20 min at 550 K. (c) $1\times10^{-7}$ mbar $O_2$ for 20 min at 550 K. Neither holes (in a,b) nor islands (in c) are observed, indicating that under the given conditions Pt is necessary for the extraction of lattice oxygen and for oxygen splitting, spillover, and island growth. The bright line passing near the center of (b) is a domain boundary in the ($\sqrt{2} \times \sqrt{2}$) R45° reconstruction [2].

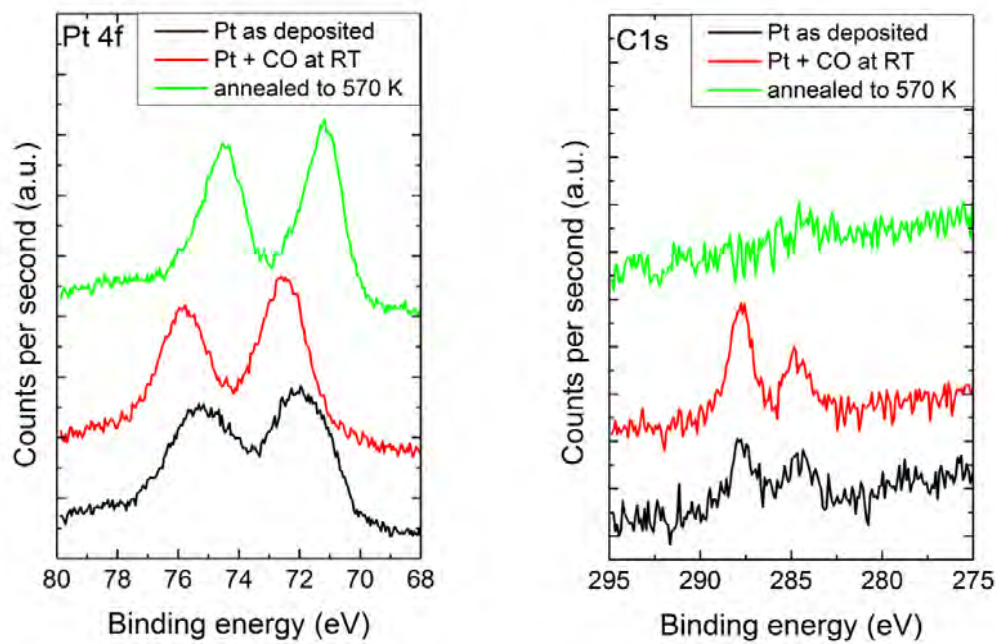

**Figure S 3: Effect of CO on Pt adatoms and clusters.** XPS spectra of the Pt 4$f$ and C 1$s$ regions for three different preparations of Pt/$Fe_3O_4$(001). Immediately after Pt deposition a broad Pt 4$f$ peak is observed. This sample comprises both Pt adatoms and small clusters. The adsorption of CO from the residual gas is already observed in the C 1$s$ region at 288 eV, as well as a small amount of C at 284.5 eV. Exposing the system to CO leads to an increase in the CO peak at 288 eV and a shift of the peak in the Pt 4$f$ spectrum to higher binding energy. After annealing to 570 K in UHV the CO desorbs resulting in a flat C1$s$ region, while the Pt 4$f$ peaks shift to a binding energy consistent with metallic Pt.

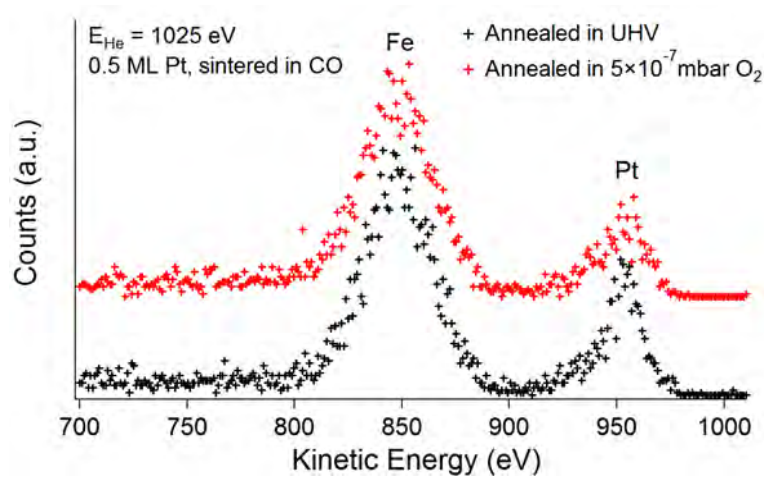

**Figure S 4: Checking for encapsulation.** Low-energy ion scattering spectra (He$^+$, incident energy 1025 eV) for two preparations of Pt/Fe$_3$O$_4$(001). The peaks at ≈850eV and ≈950eV kinetic energy correspond to backscattering from Fe and Pt, respectively. After annealing the sample with Pt clusters at 610 K in UHV for 20 min (black spectrum), no additional Fe$_3$O$_4$ islands are observed in STM images. After annealing at 610 K in 5×10$^{-7}$ mbar O$_2$ for 60 min (red spectrum), the average cluster size has increased and islands are found adjacent to almost all clusters. The Pt peak intensity after annealing in O$_2$ is smaller. However, if the clusters were fully overgrown by iron oxide, a larger difference in intensity would be expected[3]. We attribute the decrease in Pt intensity to coarsening of the Pt particles.

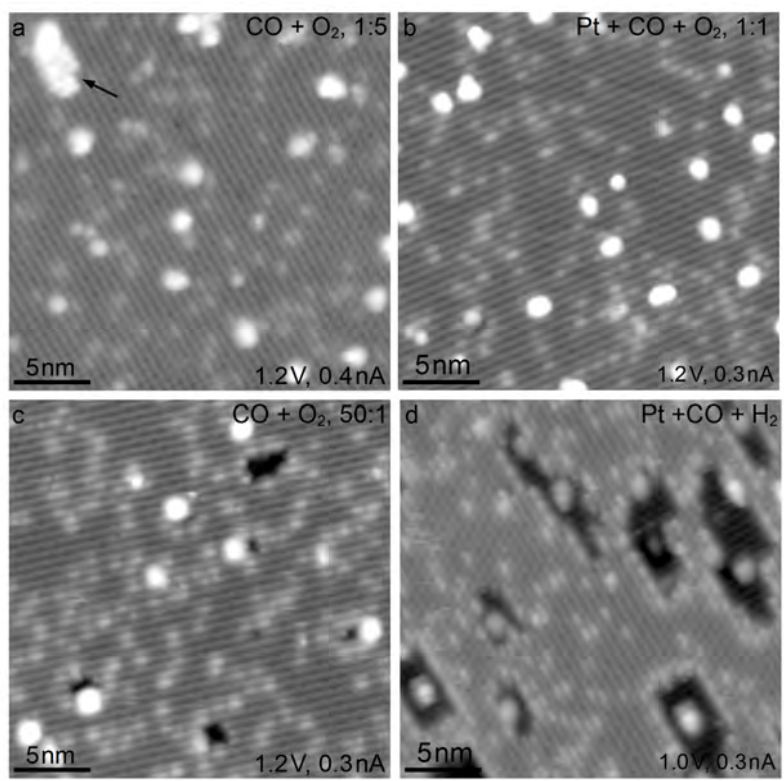

**Figure S 5: Co-dosing CO, O$_2$ and H$_2$ on the Pt$_{1-6}$/Fe$_3$O$_4$ model catalyst at 550 K:** (a) STM image acquired following exposure of the Pt$_{1-6}$/Fe$_3$O$_4$(001) sample to a 1:5 ratio of CO (1×10$^{-7}$ mbar) and O$_2$ (5×10$^{-7}$ mbar). Small Fe$_3$O$_4$(001) islands are observed, suggesting that the island-growth mechanism is not hindered by the presence of CO in the atmosphere. (b) STM image acquired following exposure of the Pt$_{1-6}$/Fe$_3$O$_4$(001) sample to a 1:1 ratio of CO and O$_2$ (1×10$^{-7}$ mbar each) for 20 minutes at 550 K. The resulting surface largely resembles the initial state (c) STM image acquired following exposure of the Pt$_{1-6}$/Fe$_3$O$_4$(001) sample to a 50:1 ratio of CO (1×10$^{-7}$ mbar) and O$_2$ (2×10$^{-9}$ mbar) for 60 min at 550 K. The observation of small holes suggests that extraction of surface O is not inhibited by the presence of low pressures of O$_2$. (d) STM image acquired following exposure of the Pt$_{1-6}$/Fe$_3$O$_4$(001) sample to a 1:1 ratio of CO and H$_2$ (1×10$^{-7}$ mbar each) for 20 minutes at 550 K. Holes in the Fe$_3$O$_4$(001) terrace up to 3 layers deep are observed.